\documentclass[conference]{IEEEtran}
\IEEEoverridecommandlockouts
\usepackage{cite}
\usepackage{amsmath,amssymb,amsfonts}
\usepackage{algorithmic}
\usepackage{graphicx}
\usepackage{textcomp}
\usepackage{xcolor}
\usepackage{hyperref}
\usepackage{booktabs} 
\def\BibTeX{{\rm B\kern-.05em{\sc i\kern-.025em b}\kern-.08em
    T\kern-.1667em\lower.7ex\hbox{E}\kern-.125emX}}
\begin{document}

\title{A Novel Approach to Network Traffic Analysis: the HERA tool\\
}

\author{\IEEEauthorblockN{Daniela Pinto$^*$, Ivone Amorim$^\dagger$, Eva Maia$^*$, Isabel Praça$^*$}
\IEEEauthorblockA{\textit{Research Group on Intelligent Engineering and Computing for Advanced Innovation and Development (GECAD),} \\
\textit{Porto School of Engineering, Polytechnic of
Porto (ISEP-IPP), 4200-072 Porto, Portugal$^*$}\\
\textit{PORTIC – Porto Research, Technology \& Innovation Center, Polytechnic of Porto (IPP), 4200-374 Porto, Portugal$^\dagger$} \\
dapsp@isep.ipp.pt, ivone.amorim@sc.ipp.pt, egm@isep.ipp.pt, icp@isep.ipp.pt}}

\maketitle

\begin{abstract}
Cybersecurity threats highlight the need for robust network intrusion detection systems to identify malicious behaviour. These systems rely heavily on large datasets to train machine learning models capable of detecting patterns and predicting threats. In the past two decades, researchers have produced a multitude of datasets, however, some widely utilised recent datasets generated with CICFlowMeter contain inaccuracies. These result in flow generation and feature extraction inconsistencies, leading to skewed results and reduced system effectiveness. Other tools in this context lack ease of use, customizable feature sets, and flow labelling options. In this work, we introduce HERA, a new open-source tool that generates flow files and labelled or unlabelled datasets with user-defined features. Validated and tested with the UNSW-NB15 dataset, HERA demonstrated accurate flow and label generation.
\end{abstract}


\section{Introduction}

Cybersecurity is essential in modern society as network attacks and threats continue to increase. Network Traffic Analysis (NTA) is a critical component of cybersecurity by monitoring and analysing data transmitted in computer networks, giving administrators insights to optimize performance~\cite{ALQUDAH2020911}. While NTA tools offer visibility into network activities to help detect security threats, they do not inherently detect them. On the other hand, Network Intrusion Detection Systems (NIDS) are specifically designed for security, identifying malicious activity or unauthorized access attempts~\cite{macia-fernandez_ugr16_2018}. These systems scan network packets in real-time to identify patterns or anomalies that may indicate a security breach, triggering alerts and automated responses, such as notifying administrators or blocking suspicious traffic~\cite{LIAO201316}. NIDS operate as a threat-detecting system using either signature-based or anomaly-based methods~\cite{LATA2022100134}. Signature-based NIDS identify threats by comparing network traffic to known malicious patterns, being effective against known threats but vulnerable to zero-day attacks, polymorphic malware, or any new unidentified threats. In contrast, anomaly-based NIDS detect deviations from typical network behaviour, enabling them to identify new attacks. However, as they flag deviations as an anomaly, they can be prone to false positives.

Artificial Intelligence (AI) is being increasingly utilised in these systems to identify potential threats more effectively~\cite{THAKKAR2020636}. Historical data is crucial, allowing models to learn and adapt to emerging threats, giving it a clear advantage over traditional signature-based methods. Furthermore, Machine Learning (ML) can detect subtle variations in network traffic, improving accuracy and reducing false positives, provided high-quality datasets are used~\cite{SOWMYA2023100827}.

In the context of NIDS, researchers actively contribute by creating new datasets and tools to facilitate data collection. CICFlowMeter is widely used for generating flow data but has issues that lead to inaccurate datasets, such as CIC-IDS2017~\cite{sharafaldin_toward_2018}, which suffers from incorrect flow descriptions, incoherent timestamps, duplicate data, and omitted attacks~\cite{lanvin_errors_2023}. Despite these flaws and expectation that others created with the same tool would exhibit similar problems~\cite{liu_error_2022}, recent datasets such as CRCDDoS2022~\cite{hadi_developing_2022}, 5GC PFCP~\cite{amponis_5g_2023}, and CIPMAIDS2023-1~\cite{ali_effective_2023}, were still created using CICFlowMeter. Since ML models rely on high-quality datasets for accurate threat predictions, there is a critical need for reliable tools to generate Network Intrusion Detection (NID) datasets~\cite{liu_error_2022}. Additionally, the inclusion of appropriate features in these datasets is equally important. ML models depend on the quality and diversity of datasets and their features for accurate event classification. However, without standardization, features are often chosen based on researchers' domain knowledge, leading to datasets with many irrelevant features that reduce ML efficiency. While there are ongoing efforts to standardize feature selection~\cite{sarhan_evaluating_2022}, this gap still requires researchers to perform feature selection.

Considering this, there is a shortage of tools for easily generating NID datasets, and widely used tools lack essential features such as ease of use, customizable feature sets, or options for labelling generated flows. This paper introduces HERA, a tool designed to streamline dataset creation, allowing users to select relevant features and providing a labelling solution during dataset generation. The main contributions of this work are:

\begin{enumerate}
    \item Presentation of the HERA tool, that allows the generation of flow-based datasets and traffic statistics from Packet Capture (PCAP) files.
    \item HERA testing and validation with anomaly detection using ML models and the files from the dataset UNSW-NB15.
\end{enumerate}

The rest of the paper is structured as follows. Section 2 presents the concepts related to the developed work. Section 3 summarizes recent research contributions on tools for dataset creation. Section 4 describes the implemented tool. Section 5 elaborates on HERA’s validation through experiments and analysis of results. Finally, Section 6 concludes and outlines future work planned for HERA.

\section{Background}


This section revises key concepts to ensure a clear understanding of the ideas discussed in the paper. It covers the use of packet and flow information in datasets, highlighting the advantages and disadvantages of each approach. Additionally, we provide an overview of tools commonly used in dataset creation, categorized by their functions: Generator, Capturer, Flow Exporter, Feature Extractor, Flow Collector, and Classifier~\cite{pinto_hera_2024}. Finally, we outline the possible approaches for dataset creation.


NID datasets can be composed of either packet or flow information. Packet-based datasets provide a detailed view of each packet, allowing deep inspection of network protocols. However, they are resource-intensive, requiring more storage and computational power, making them less scalable. In contrast, flow-based datasets aggregate packets into flows, reducing the amount of data processed and making them more efficient, with lower storage requirements~\cite{wilailux_novel_2021}. Flow data offers a broader view of network traffic, making it well-suited for detecting malicious patterns. While packet-based analysis is ideal for scenarios needing high granularity, flow-based analysis is better suited for larger systems and is especially useful for NIDS~\cite{vormayr_why_2020}.


\emph{Generators} create synthetic traffic at the network, packet, and flow levels. These tools are essential to simulate traffic in scenarios where generating real network activity is impractical or impossible. A key advantage of this approach is the ability to fully control the generated traffic, which is particularly useful for creating datasets for analysis or testing. A tool with this function is Ostinato\footnote{\href{https://github.com/pstavirs/ostinato}{https://github.com/pstavirs/ostinato}}.


The most commonly used tool in dataset creation is the \emph{Capturer}. This tool captures packets from communication sessions between devices on a network and saves them in files, namely in the PCAP format. Researchers use these files to extract features, generate flows for easier analysis, or replay the captured packets for further study. These tools support post-capture analysis of communications and provide data for creating NID datasets. A commonly used tool for this function is tcpdump\footnote{\url{https://www.tcpdump.org/}}. 


To obtain flow data for easier-to-manage datasets, \emph{Flow Exporters} aggregate network packets into flows and export them to \emph{Flow Collectors}. These tools can capture network packets directly from interfaces or use network capture files, such as PCAPs, obtained by \emph{Capturers}, to generate flows. \emph{Flow Exporters} group packets based on properties defined by the tool, which can vary depending on the protocol and version used. They may support protocols such as NetFlow, jFlow, sFlow, or IPFIX, depending on compatibility~\cite{vormayr_why_2020}. A commonly used tool is CICFlowMeter\footnote{\url{https://www.unb.ca/cic/research/applications.html}}. 


\emph{Feature Extraction} is commonly used in generating NID datasets and is often linked with \emph{Flow Exporters} although it is not essential for its primary function and can also be applied to packet data alone. It involves transforming raw data into features that contain relevant information for analysis. These features are essential for NIDS. For flow data, typical features include flow identifiers, packet counts, flow duration, protocol types, flow direction, and other details based on the dataset author’s requirements. A tool that mainly tries to execute this function is LiteEX~\cite{swarnkar_liteex_2023}. 


\emph{Flow Collectors} receive flows from \emph{Flow Exporters} and handle their storage and preliminary processing. While not as commonly used as other tools, they offer the advantage of aggregating flows from multiple \emph{Flow Exporters} within a network, making analysis more efficient. Depending on user needs and network data volume, flows are stored in databases or file systems. The main advantage of using \emph{Flow Collectors} is their capacity to handle data from large networks, enabling comprehensive monitoring and improving data quality through processes such as duplicate removal. They are particularly effective when integrated with other network monitoring tools. A tool commonly used as a \emph{Flow Collector} is  SiLK\footnote{\href{https://tools.netsa.cert.org/silk/}{https://tools.netsa.cert.org/silk/}}.


\emph{Classifiers} categorize different types of traffic, such as distinguishing between sources (e.g., email, voice) or identifying benign versus malicious traffic. These tools improve security monitoring and threat detection by recognizing attack patterns within datasets, making them valuable in network environments. An example of this type of tool is FlowTransformer~\cite{zhao_flow_2021}.

\begin{figure}[!htbp]
    \centering
    \includegraphics[width=0.8\linewidth]{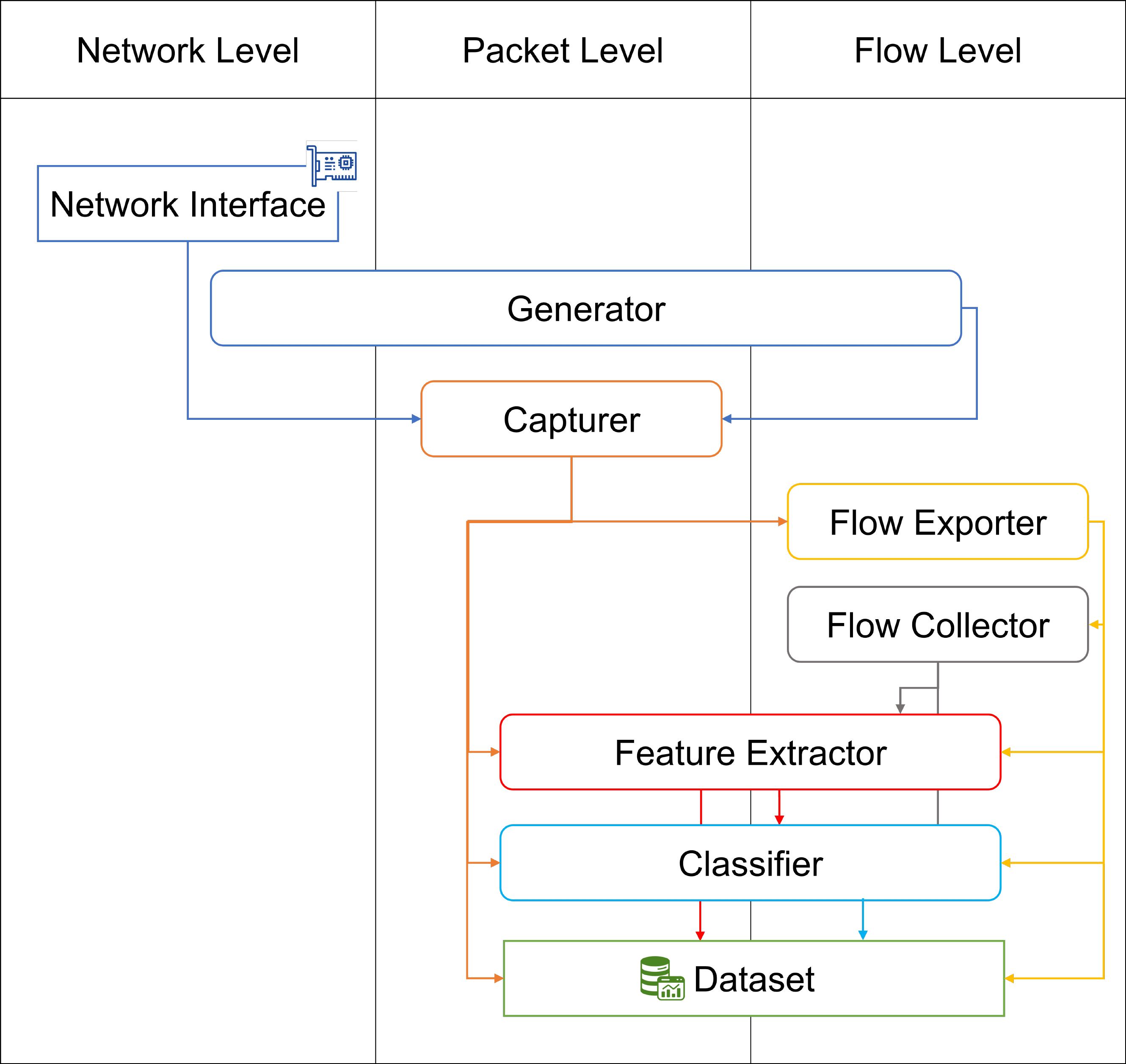}
    \caption{Dataset creation process.}
    \label{fig:classdia}
\end{figure}

Fig.~\ref{fig:classdia} illustrates the interactions among these tools in dataset generation. 

Datasets can be created through different paths depicted in the figure. One option is to create a dataset from a simple PCAP file, generated by a \emph{Capturer}, either collected from a network interface or created using a synthetic traffic \emph{Generator}. If the dataset remains at the packet level, it can be enhanced with a \emph{Feature Extractor} or \emph{Classifier} to add more features and labels. Alternatively, for a flow-based dataset, the captured traffic can be processed through a \emph{Flow Exporter} to generate a flow-only dataset. In larger networks, a \emph{Flow Collector} can aggregate flows from multiple \emph{Flow Exporters} before passing the data to a \emph{Feature Extractor} and/or \emph{Classifier}, thereby generating a more complex flow-based dataset.

After revising the key concepts, the next section will examine the existing work on this topic and the research conducted by the scientific community.

\section{Related Work}

Research on NID datasets has been extensive, with a primary focus on dataset generation. Recently, however, more tools have been developed to facilitate this process, especially tools for generating flow data. CICFlowMeter, developed by the Canadian Institute for Cybersecurity (CIC), is among the most widely used tools for generating flow data. It serves as a Flow Exporter, aggregating flows bidirectionally from packets and is capable of capturing packets directly or converting PCAP files into flows. CICFlowMeter is not only used by CIC researchers to produce datasets~\cite{habibi_lashkari_characterization_2016, lashkari_characterization_2017} but also by other researchers~\cite{hadi_developing_2022, ali_effective_2023, amponis_5g_2023}. However, several researchers have identified errors in this tool. In 2021, Engelen et al.~\cite{engelen_troubleshooting_2021} revealed that CICFlowMeter incorrectly marked a Transmission Control Protocol (TCP) connection as closed upon detecting a FIN flag, even if the protocol termination was incomplete. They also found a timing-related flaw that altered the source and destination in the generated flows and noted that CICFlowMeter did not consider the TCP Reset flag as a valid method for closing a TCP flow. In 2022, Liu et al.~\cite{liu_error_2022} demonstrated further incorrect implementations in feature creation, resulting in mislabeled flows. Additionally, in 2021, Rosay et al.~\cite{rosay_cic-ids2017_2021} observed feature duplication, feature miscalculation, and incorrect protocol detection. These authors, along with Lanvin et al.~\cite{lanvin_errors_2023}, have proposed corrections to mitigate CICFlowMeter’s issues, with Rosay et al.~\cite{rosay_cic-ids2017_2021} developing a new tool, LycoSTand, for feature extraction. 

LycoSTand functions similarly to CICFlowMeter by reading PCAP files, processing packets into flows, and calculating 82 flow-based features, equivalent to those in CICFlowMeter, generating a CSV file for each PCAP file processed. For feature extraction, LycoSTand initializes all features to 0 at the start of a flow and incrementally updates them as each new packet is added, using a method that avoids the computational errors seen in CICFlowMeter. A main correction in LycoSTand is in the TCP session terminations, addressing a major issue present in CICFlowMeter.

ICSFlowGenerator, introduced by Dehlaghi-Ghadim et al.~\cite{dehlaghi-ghadim_anomaly_2023}, combines the functionalities of a Flow Exporter and Feature Extractor. It processes PCAP files, forming flows based on a specified time interval, and outputs a CSV file containing flow-based features, categorized into flow, general, or TCP-specific features. ICSFlowGenerator also includes an automatic labelling system that uses Injection Timing and Network Security Tools to mark traffic as benign or malicious. This tool offers two labelling methods for identifying malicious flows: one based on attack history logs and another by using pre-trained models to predict anomalies by analysing flows with predicated labels.

Finally, Swarnkar et al.~\cite{swarnkar_liteex_2023} developed LiteEx, a tool that focuses mainly on feature extraction and is intended to be lightweight and flexible, providing a Graphical User Interface (GUI). To diminish the computational cost of flow formation, the authors built the tool to construct the flow temporarily and extract only the features. When testing it, the authors found it to extract the features more efficiently than similar tools, but it became less efficient for larger datasets.

The research community's efforts toward developing tools that facilitate data collection for integration into NIDS datasets have been steadily increasing, reflecting a growing need for reliable tools capable of generating high-quality datasets with ease. However, the solutions presented thus far have limitations~\cite{swarnkar_liteex_2023}. For instance, they restrict feature extraction, not allowing for a customizable feature set (e.g. CICFlowMeter, LycoSTand, ICSFlowGenerator and LiteEX), struggle with efficiently processing large volumes of data (e.g. LiteEX), are not publicly available (e.g. LiteEX) and, in the case of CICFlowMeter generate data with errors. As a result, a streamlined process for obtaining datasets for NID research does not currently exist, leading to the need for a new tool.

\section{HERA}

In this section, we present our novel tool, named Holistic nEtwork featuRes Aggregator (HERA). This tool allows users to produce datasets, with or without labels, for use in NID research.
HERA provides an easy and streamlined alternative to dataset creation, motivated by the issues identified with the popular tool CICFlowMeter and the errors found in the datasets produced using it.

HERA follows the typical approach when developing a flow-based dataset, as illustrated in Fig.~\ref{fig:classdia}, which begins by involving a \emph{Capturer} listening on a network interface to capture traffic in a PCAP file. To note, this is done externally to the tool, as HERA receives the PCAP files as input. Subsequently, this is processed by a \emph{Flow Exporter}, Argus, to generate flows which perform \emph{Feature Extraction}. The resulting flow data and features are then stored as a dataset in CSV format.






\subsection{HERA Workflow Overview}

\begin{figure}[!htbp]
    \centering
    \includegraphics[width=0.9\linewidth]{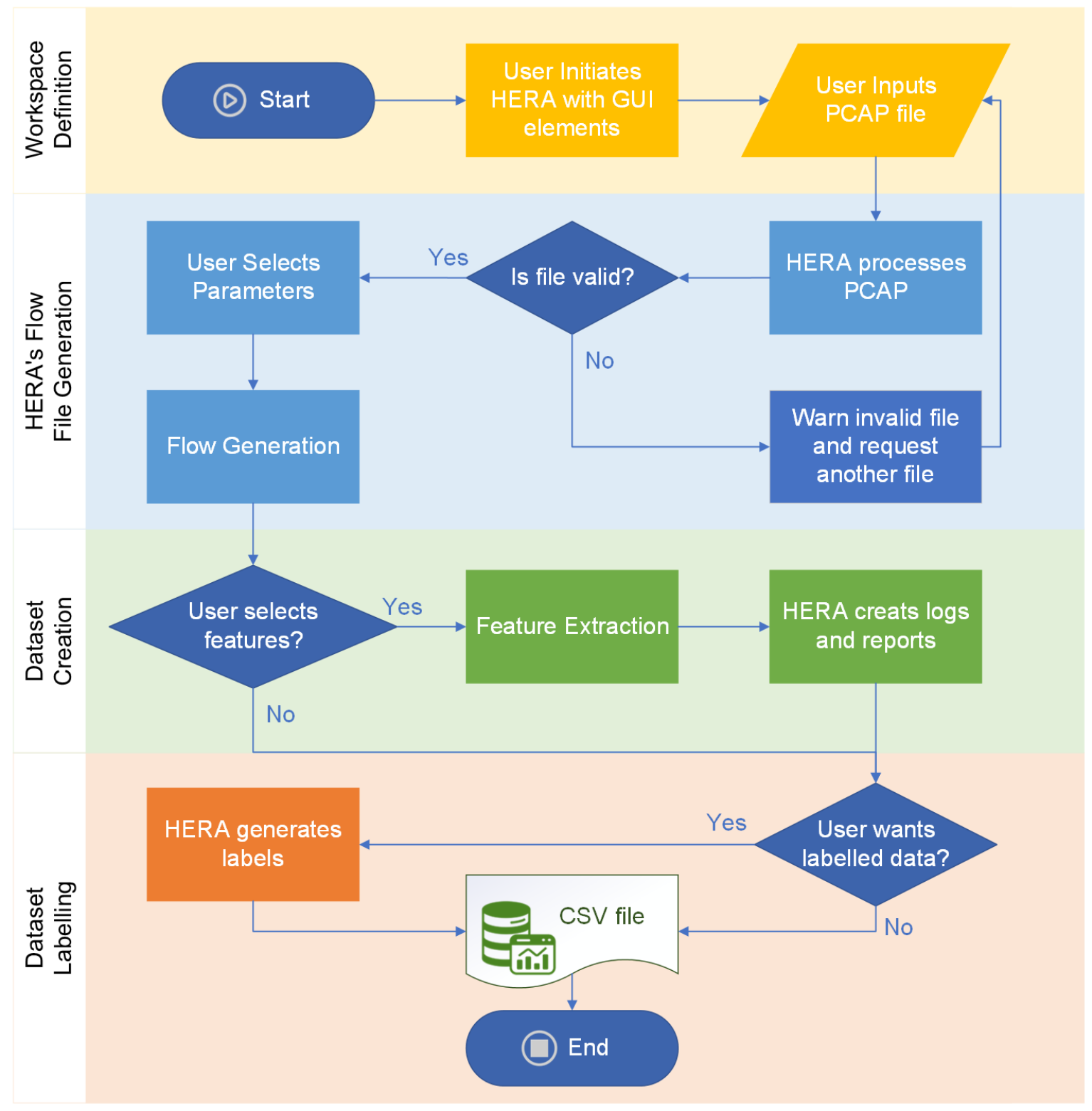}

    \caption{HERA's flowchart with the main components identified.}
    \label{fig:tool_flowchart}
\end{figure}

HERA leverages the Flow Exporter Argus to obtain flows and extract features. This choice allows our tool to use either PCAP files or HERA flow files, which consist of argus data, to generate datasets. These flow files can be created by customising fields like flow interval and the features that will integrate the datasets. Additionally, a labelled version of the dataset can be generated with the use of a ground truth file. Moreover, HERA was designed to be easy to use and, as such, is divided into four components:
 \emph{Workspace Definition}, \emph{HERA's Flow File Generation}, \emph{Dataset Creation}, and \emph{Dataset Labelling}. The last three components produce specific outputs (Fig.~\ref{fig:tool_flowchart}).

Initially, the users import the necessary libraries, and during the \emph{Workspace Definition} phase, they are prompted to specify the path locations for the PCAP files they intend to convert into flows, as well as the destination for the generated HERA and CSV files. In the subsequent component, which pertains to HERA’s flow file generation (optional if the user has already obtained these files), the user selects arguments to add information to the flows. This is accomplished using subprocesses with Argus, and the resulting data is stored in a \textit{.hera} file. The user also defines the interval between flows before the generation of the HERA flow files. In the \emph{Dataset Creation} component, HERA utilises the generated file to create a dataset based on a set of user-selected features, allowing the user to choose the client they wish to use (ra or racluster). Additionally, the user decides whether to retain or discard Argus management flows in the final CSV file. Finally, in the \emph{Dataset Labelling} component, after specifying the location of the ground truth file, HERA produces the same dataset as previously, but with an additional column containing labels.


In the following sections, we will provide a detailed description of the three output-generating components. 

\subsection{HERA's Flow File Generation}

As previously mentioned, for the Flow Exporter functionality of HERA, the well-known tool Argus was selected. 
Argus is a versatile tool that allows for capturing and reporting flow data and integrating real-time network traffic analysis with other tools. With over 30 years of continuous contributions, Argus has been recognized for its robust functionalities and is recommended by the European Union Agency for Cybersecurity (ENISA)~\cite{enisa}. Consequently, Argus has been utilised in multiple projects as a Flow Exporter, generating data for datasets such as UNSW-NB15\footnote{\href{https://research.unsw.edu.au/projects/unsw-nb15-dataset}{https://research.unsw.edu.au/projects/unsw-nb15-dataset}} and BoT-IoT\footnote{\href{https://research.unsw.edu.au/projects/bot-iot-dataset}{https://research.unsw.edu.au/projects/bot-iot-dataset}}.

HERA uses several programs from Argus, including argus, ra, racluster, and racount. The argus program is specifically used in this component of HERA, \emph{Flow File Generation}, via a subprocess that facilitates the formation of flows. While racount is used by HERA in this step to return statistics of the generated flows in the form of text files.

In this component, the user starts by selecting arguments from a list to determine what information to include in the flow file. Next, they specify the interval for the flows, with a default value of 60 and only positive numbers being accepted. Finally, the user initiates HERA to generate the flow file, which is then saved as a \emph{.hera} file.

\subsection{Dataset Creation}

The third component of HERA produces the unlabelled dataset by allowing the user to select from 130 possible features, two of which are calculated by HERA. These two features calculate the number of connections with the same service and source or destination address. Furthermore, by selecting ra or racluster, the user can choose to obtain the original flows generated in the previous component or an aggregated version. 
Ra is a client that will present all the flows generated in the previous step, while racluster is useful for larger datasets, allowing for efficient processing with summarized information. Racluster aggregates flows based on the flow-defining tuples, returning a smaller dataset with the same flow configuration settings as ra would use. Furthermore, the user can select from five predefined feature sets: one containing all features, three sets of features present in popular datasets, UNSW-NB15, BoT-IoT, and CIC-IDS2017, and a default set.

The default set includes the following features: ``FlowID'' (flow identifier), ``rank'' (the ordinal value of the flow record, representing its sequence number), ``stime'' and ``ltime'' (record's start and last time), ``sport'' and ``dport'' (source and destination port number), ``saddr'' and ``daddr'' (source and destination Internet Protocol (IP) address), ``proto'' (identification of the transaction protocol), ``bytes'', ``sbytes'' and ``dbytes'' (total, source to destination, and destination to source transaction bytes), ``pkts'', ``spkts'' and ``dpkts'' (total, source to destination, and destination to source packet count), ``dur'' (total duration of the flow record), ``runtime'' and ``idle'' (total active flow runtime and time since the last packet activity), ``flgs'' (flow state flags), ``tcpopt'' (flag for the TCP connection state), and ``Ssaddr'' and ``Sdaddr'' (calculated features for the number of connections with the same service and source or destination address, respectively).

It is important to note that the features ``rank'', ``stime'', ``ltime'', ``proto'', ``saddr'', ``daddr'', ``sport'' and ``dport'' are always generated. The ``FlowID'' is created by merging destination and source addresses, ports, and protocols, separated by hyphens. Finally, by using a subprocess with ra or racluster, and racount, HERA extracts the dataset to a CSV file and statistics from the flows into a text file.

\subsection{Dataset Labelling}

Regarding the final component, after the user indicates the location of the ground truth file, HERA uses this information to label the flows with a specified label or classify them as benign if the particular flow is not identified in the ground truth file. For this step, the collected traffic must be captured in a controlled environment where the authors can accurately describe the network activity. Consequently, the provided ground truth file must include headers for ``StartTime'', ``LastTime'', ``Proto'', ``SrcAddr'', ``Sport'', ``DstAddr'', ``Dport'', and ``Label'', corresponding to the start and end times of the attack, the protocol used, the source and destination IP addresses and ports, and the label to be applied. The only mandatory field is the label identifying the attack, allowing flexibility when some information is unknown, as only some fields may be sufficient to distinguish an attack from background traffic.

Afterwards, HERA stores this information and uses it to label the previously generated unlabelled dataset, iterating through the flows to check for matches with the ground truth file information. To enhance efficiency, HERA verifies time ranges in the unlabelled dataset to determine if all rows in the ground truth file need to be checked for the presence of attacks. This is because a dataset can consist of several CSV files, while the entire ground truth can be a single file.

To finalize, HERA provides, alongside the labelled dataset, a text file summarizing the total number of flows that identify malicious or benign traffic.




\section{Validation, Experiments and Results}

To verify that HERA produces flows correctly, PCAP files from the UNSW-NB15 dataset were used, as this dataset was generated with the Flow Exporter Argus.

UNSW-NB15 is a dataset created to address the limitations found in older datasets such as KDDCUP'99 and NSL-KDD. It was generated in a synthetic environment using the IXIA tool testbed over two days, with traffic collected for 16 hours on the first day and 15 hours on the second. The dataset includes attacks generated every second on the first day and ten attacks per second on the second day. The generated flows and features were obtained using BRO-IDS/Zeek\footnote{\href{https://zeek.org/}{https://zeek.org/}}, a tool with Intrusion Detection System (IDS) like capabilities whose logs have been used by researchers to obtain data for datasets, and Argus, saved in a database by matching the features of the two and running algorithms to obtain additional features. The result is a dataset saved in a CSV file with benign and malicious traffic, composed of Fuzzers, Analysis attacks, Backdoors, Denial of Service (DoS), Exploits, Generic attacks, Reconnaissance, Shellcode and Worms.

The advantage of using this dataset is the extensive documentation provided by the researchers, which includes the flow files obtained from Argus with unlabeled CSV files, the PCAP files used, and detailed documentation on features and the ground truth file. For the tests, only the files relevant to the second day were used, as the PCAP files from the first day contained repetitions, making it impossible to obtain comparable results with HERA. In the first test, UNSW-NB15's Argus files were compared to those obtained by HERA, resulting in two key observations. Firstly, the flow files from both sources had the same number of generated flows. However, a discrepancy of an additional 23,204 flows was observed in the researchers' unlabelled CSV files containing only Argus data, likely due to flow and feature processing before dataset creation. Furthermore, the final labelled dataset contained 339,727 additional flows compared to those obtained using only Argus, possibly added exclusively by Bro-IDS/Zeek.

Additional tests using ML techniques were conducted to evaluate the quality of the generated datasets (both HERA's version and those provided by the researchers) using supervised and unsupervised learning. The next sections present the validation of HERA with the results obtained using these experiments.

\subsection{Supervised Learning}

Supervised Learning techniques were used to evaluate the quality of the generated datasets using three models: Random Forest, Light Gradient Boosting Machine, and Extreme Gradient Boosting. These models were chosen for their performance and efficiency on large datasets. Initially, feature importance analysis was conducted using Random Forest's ``feature\_importance\_'' attribute in scikit-learn\footnote{\href{https://scikit-learn.org/stable/}{https://scikit-learn.org/stable/}}. This analysis allowed us to select the most relevant features for training the models: ``sbytes'', ``dbytes'', ``sttl'', ``dttl'', ``smeansz'', ``dmeansz'' and ``synack''. Afterwards, these models were tested with different parameters to obtain the best results. These are presented in Table~\ref{tab:unswnb15_ml_results}.

\begin{table}[!htbp]
    \centering
    \caption{Results of algorithms applied to the UNSW-NB15 dataset.} \label{tab:unswnb15_ml_results} 
    \begin{tabular}{ ccccc }
        \toprule
        \textbf{Algorithm} & \textbf{Precision} & \textbf{Recall} & \textbf{F1-score} & \textbf{Accuracy} \\
        \midrule
        \textbf{RandomForestClassifier} & \textbf{0.98} & \textbf{0.98} & \textbf{0.98} & \textbf{0.9894} \\ \hline
        LGBMClassifier & 0.97 & 0.97 & 0.97 & 0.9890 \\ \hline
        XGBClassifier & 0.97 & 0.97 & 0.97 & 0.9888 \\ 
        \bottomrule
    \end{tabular}
\end{table}  

The best result was obtained with the Random Forest algorithm, which was then selected to train the model. The labelled version of the UNSW-NB15 dataset, primarily using data from the second day, was employed to train a model for classifying the unlabelled datasets generated by HERA, named ``HERA's CSV'', and another consisting of the researchers' unlabelled dataset using only Argus data, named ``Researcher's CSV''. To note, the labelled version of UNSW-NB15 is here referred to as UNSW-NB15\_3 + UNSW-NB15\_4 as these datasets are divided into four files with the last two being composed exclusively of second-day traffic, even if not the entire day. Additionally, due to the presence of NaN values in the dataset generated by HERA, an imputer using the mean strategy was applied to replace these values. Since the models were used to evaluate unlabelled datasets, the proportion of malicious traffic identified by the models in the datasets was compared to assess whether HERA produced accurate results.

\begin{figure}[htbp]
\centerline{\includegraphics[width=0.8\linewidth]{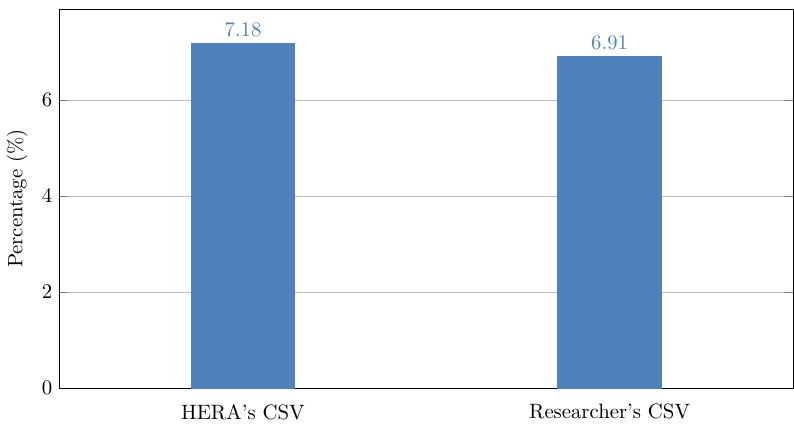}}
\caption{Percentage of malicious traffic in the different versions of the dataset.}
\label{fig:unswnb15_per_mal}
\end{figure}

The original dataset, which combines the UNSW-NB15\_3 and UNSW-NB15\_4 CSV files, contained 21.61\% malicious traffic. When this dataset was used to train the model, it produced favourable results. However, when the model was used to classify datasets generated by HERA and the Researcher’s CSV files, the results were similar to each other but significantly lower than those of the original dataset, with 7.18\% and 6.91\% malicious traffic respectively, as shown in Fig.~\ref{fig:unswnb15_per_mal}. The similarity in results between the HERA and Researcher’s datasets supports the validity of HERA, as both were generated using the same Flow Exporter. The discrepancy between the original dataset and these two is likely due to differences in the total number of flows, as previously mentioned.

\subsection{Unsupervised Learning}

For unsupervised learning, K-means was used with the same features as previously with a cluster of two to identify benign or malicious traffic. Using this strategy the following sets of datasets obtained the silhouette scores presented in Table~\ref{tab:silun}. Using this model, it is demonstrated that HERA obtains the most well-defined clusters, demonstrating an improvement using this tool.

\begin{table}[!htbp]
    \centering
    \caption{Silhouette Scores for unsupervised K-means (k=2).} \label{tab:silun} 
    \begin{tabular}{ cc }
        \toprule
        \textbf{Model} & \textbf{Silhouette Score} \\
        \midrule
        UNSW-NB15\_3 + UNSW-NB15\_4 & 0.5061 \\ \hline
        \textbf{HERA's CSV} & \textbf{0.7258}  \\ \hline
        Researcher's CSV & 0.6879  \\ 
        \bottomrule
    \end{tabular}
\end{table}  

\begin{figure}[htbp]
\centerline{\includegraphics[width=0.8\linewidth]{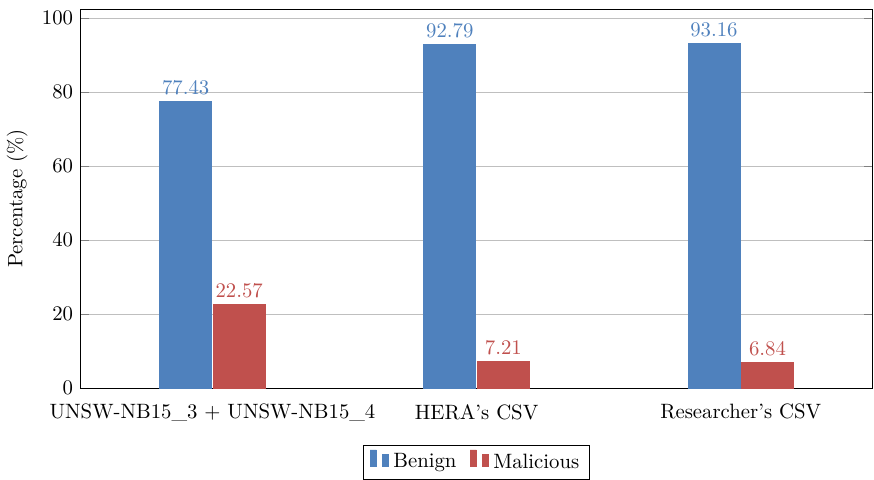}}
\caption{Percentage of malicious traffic in different dataset versions (k=2).}
\label{fig:unswnb15_unsu_2}
\end{figure}

Fig.~\ref{fig:unswnb15_unsu_2} demonstrates the obtained results using unsupervised learning regarding the percentages of identified malicious flows. As previously mentioned, the combination of UNSW-NB15\_3 and UNSW-NB15\_4 has 21.61\% of malicious traffic. With the unsupervised model, the combination of these two files obtained a similar amount of malicious traffic at 22.57\%. Additionally, for HERA’s version and the Researcher’s CSV, both the supervised and unsupervised models obtained similar results with 7.18\% and 7.21\% for the first and 6.91\% and 6.84\% for the second. This is again explainable by the difference in flows in the original UNSW-NB15 dataset with more flows, in specific, malicious, added by BRO-IDS/Zeek.

\subsection{Labelling the Dataset}

As a final test, using the labelling feature of HERA, the dataset was labelled using the ground truth the researchers provided. This confirmed the amount of malicious flows in the dataset obtaining 6.56\% malicious traffic a close value to that obtained with ML. Table~\ref{tab:label_unsw} breaks down the total amount of flows identified in the dataset generated with HERA and the dataset containing Argus and BRO-IDS/Zeek data. As it is observable, with BRO-IDS/Zeek data, UNSW-NB15 contains 10 times more of the total amount of Generic attack traffic with generally all attack types and benign traffic having higher amounts of flows.

\begin{table}[!htbp]
    \centering
    \caption{Results of the labelling process.} \label{tab:label_unsw} 
    \begin{tabular}{ ccc }
        \toprule
        \textbf{Type} & \textbf{HERA's CSV} & \textbf{UNSW-NB15\_3 + UNSW-NB15\_4} \\
        \midrule
        Benign & 726.153 & 893.726 \\ \hline
        Exploits & 16.157 & 28.013 \\ \hline
        Fuzzers & 12.060 & 14.527 \\ \hline
        Generic & 11.468 & 180.076 \\ \hline
        Reconnaissance & 7.366 & 9.112 \\ \hline
        DoS & 2.294 & 10.549 \\ \hline
        Shellcode & 953 & 964 \\ \hline
        Analysis & 309 & 1.543 \\ \hline
        Backdoor & 232 & 1.425 \\ \hline
        Worms & 104 & 110 \\ \hline \hline
        \textbf{Total} & 777.096 & 1.140.045 \\
        \bottomrule
    \end{tabular}
\end{table} 

Finally, the models in the supervised learning were used in the labelled HERA dataset and obtained the results presented in Table~\ref{tab:unswnb15_ml_results_hera}. As expected, because the original UNSW-NB15 dataset has more malicious traffic added by BRO-IDS/Zeek, the model did not learn as accurately in the HERA version what malicious traffic was.

\begin{table}[!htbp]
    \centering
    \caption{Results of algorithms applied to the labeled HERA dataset.} \label{tab:unswnb15_ml_results_hera} 
    \begin{tabular}{ ccccc }
        \toprule
        \textbf{Algorithm} & \textbf{Precision} & \textbf{Recall} & \textbf{F1-score} & \textbf{Accuracy} \\
        \midrule
        \textbf{RandomForestClassifier} & \textbf{0.88} & \textbf{0.88} & \textbf{0.88} & \textbf{0.9842}  \\ \hline
        LGBMClassifier & 0.88  & 0.87 & 0.87 & 0.9837 \\ \hline
        XGBClassifier & 0.88 & 0.87 & 0.87 & 0.9834 \\ 
        \bottomrule
    \end{tabular}
\end{table} 

Nevertheless, it can be said that the tool successfully generated the expected number of flows from the UNSW-NB15 dataset, validating its correct functionality. Additionally, ML tests for anomaly detection confirmed that HERA produces results comparable to those obtained using the original CSV files containing only Argus information, demonstrating the tool's effectiveness.   

\section{Conclusion and Future Work}

This paper presents a solution for streamlined dataset creation with HERA, an open-source tool that generates flow files, unlabelled datasets, and labelled datasets. HERA allows users to customize parameters, select relevant features, and, with a ground truth file, easily create labelled datasets.

Compared to CICFlowMeter, HERA offers several advantages: customizable flow parameters, customizable feature sets, flow labelling, and three outputs: flow files and both labelled and unlabelled datasets.

Experiments demonstrated that HERA accurately uses the Flow Exporter Argus to generate flows, matching the flow count in the UNSW-NB15 dataset. The tool's efficiency was confirmed with ML algorithms, and its labelling component correctly reflected the proportions of malicious traffic.

In the future, we intend for HERA to include a packet-capturing component to handle the entire dataset creation process, to integrate it with other tools for real-time use, and to add anomaly detection using ML on generated datasets.

\section*{Acknowledgment}

This work was supported by the CYDERCO project, which has received funding from the European Cybersecurity Competence Centre under grant agreement 101128052. This work has also received funding from UIDB/00760/2020.


\bibliographystyle{IEEEtran}
\bibliography{refs}

\end{document}